\documentclass[10pt, conference, compsocconf]{IEEEtran}

\usepackage{graphicx}

\begin{document}
\title{Simulation Analysis of IEEE 802.15.4\\ Non-beacon Mode at Varying Data Rates}

\author{Z. Abbas, N. Javaid, M. A. Khan, S. Ahmed$^\sharp$, U. Qasim$^{\ddag}$, Z. A. Khan$^{\S}$\\

        COMSATS Institute of IT, Islamabad, Pakistan. \\
        $^\sharp$Mirpur University of Science and Technology, AJK, Pakistan.\\
        $^{\ddag}$University of Alberta, Alberta, Canada.\\
        $^{\S}$Faculty of Engineering, Dalhousie University, Halifax, Canada.
        }

\maketitle

\begin{abstract}
%\boldmath
IEEE 802.15.4 standard is designed for low power and low data rate applications with high reliability. It operates in beacon enable and non-beacon enable modes. In this work, we analyze delay, throughput, load, and end-to-end delay of non-beacon enable mode. Analysis of these parameters are performed at varying data rates. Evaluation of non beacon enabled mode is done in a 10 node network. We limit our analysis to non beacon or unslotted version because, it performs better than other. Protocol performance is examined by changing different Medium Access Control (MAC) parameters. We consider a full size MAC packet with payload size of 114 bytes. In this paper we show that maximum throughput and lowest delay is achieved at highest data rate.
\end{abstract}

\begin{IEEEkeywords}
IEEE 802.15.4, Throughput, Delay, End-to-end, Load
\end{IEEEkeywords}

\section{Introduction}
During past few years research in areas of Wireless Ad-hoc Networks and Wireless Sensor Networks (WSNs) are escalated. IEEE 802.15.4 is targeted for Wireless Body Area Networks (WBANs), which requires low power and low data rate applications. Invasive computing is term used to describe future of computing and communications [1-3]. Due to these concepts, personal and business domains are being densely populated with sensors. One area of increasing interest is the adaptation of technology to operate in and around human body. Many other potential applications like medical sensing control, wearable computing and location identification are based on Wireless Body Area Networks (WBANs).

Main aim of IEEE 802.15.4 standard is to provide a low-cost, low power and reliable protocol for wireless monitoring of patient's health. This standard defines physical layer and MAC sub layer. Three distinct frequencies bands are supported in this standard. However, 2.4 GHz band is more important. This frequency range is same as IEEE 802.11b/g and Bluetooth. IEEE 802.15.4 network supports two types of topologies, star topology and peer to peer topology. Standard supports two modes of operation, beacon enabled (slotted) and non-beacon enabled (unslotted).

Medium Access Control (MAC) protocols play an important role in overall performance of a network. In broad, they are defined in two categories contention-based and schedule-based MAC protocols. In contention-based protocols like Carrier Sense Multiple Access with Collision Avoidance (CSMA/CA), each node content to access the medium. If node finds medium busy, it reschedules transmission until medium is free. In schedule-based protocols like Time Division Multiple Access (TDMA), each node transmits data in its pre-allocated time slot.

This paper focuses on analysis of IEEE 802.15.4 standard with non-beacon enabled mode configure in a star topology. We also consider that sensor nodes are using CSMA/CA protocol. To access channel data.

%This paper is organized as, section II contains related work and motivation, section III describes overview and operation of IEEE 802.15.4. Simulation study of IEEE 802.15.4 are presented in section IV. Section V concludes the work with future work.

\section{Related Work and Motivation}
In literature, beacon enabled mode is used with slotted CSMA/CA for different network settings. In [1], performance analysis of IEEE 802.15.4 low power and low data rate wireless standard in WBANs is done. Authors consider a star topology at 2.4 GHz with up to 10 body implanted sensors. Long-term power consumption of devices is the main aim of their analysis. However, authors do not analyze their study for different data rates.

An analytical model for non-beacon enabled mode of IEEE 802.15.4 medium access control (MAC) protocol is provided in [2]. Nodes use un-slotted CSMA/CA operation for channel access and packet transmission. Two main variables that are needed for channel access algorithm are Back-off Exponent (BE) and Number of Back-offs (NB). Authors perform mathematical modeling for the evaluation statistical distribution of traffic generated by nodes. This mathematical model allows evaluating an optimum size packet so that success probability of transmission is maximize. However, authors do not analyze different MAC parameters with varying data rates.

Authors carry out an extensive analysis based on simulations and real measurements to investigate the unreliability in IEEE 802.15.4 standard in [3]. Authors find out that, with an appropriate parameter setting, it is possible to achieve desired level of reliability. Unreliability in MAC protocol is the basic aspect for evaluation of reliability for a sensor network. An extensive simulation analysis of CSMA/CA algorithm is performed by authors to regulate the channel access mechanism. A set of measurements on a real test bed is used to validate simulation results.

A Traffic-adaptive MAC protocol (TaMAC) is introduced by using traffic information of sensor nodes in [4]. TaMAC protocol is supported by a wakeup radio, which is used to support emergency and on-demand events in a reliable manner. Authors compare TaMAC with beacon-enabled IEEE 802.15.4 MAC, Wireless Sensor MAC (WiseMAC), and Sensor MAC (SMAC) protocols.

Important requirements for the design of a low-power MAC protocol for WBANs are discussed in [5]. Authors present an overview to Heartbeat Driven MAC (H-MAC), Reservation-based Dynamic TDMA (DTDMA), Preamble-Based TDMA (PB-TDMA), and Body MAC protocols, with focusing on their strengths and weaknesses. Authors analyze different power efficient mechanism in context of WBANs. At the end authors propose a novel low-power MAC protocol based on TDMA to satisfy traffic heterogeneity.

Authors in [6], examine use of IEEE 802.15.4 standard in ECG monitoring and study the effects of CSMA/CA mechanism. They analyze performance of network in terms of transmission delay, end-to-end delay, and packet delivery rate. For time critical applications, a payload size between 40 and 60 bytes is selected due to lower end-to-end delay and acceptable packet delivery rate.

In [7], authors state that IEEE 802.15.4 standard is designed as a low power and low data rate protocol with high reliability. They analyze unslotted version of protocol with maximum throughput and minimum delay. The main purpose of IEEE 802.15.4 standard is to provide low power, low cost and highly reliable protocol. Physical layer specifies three different frequency ranges, 2.4 GHz band with 16 channels, 915 MHz with 10 channels and 868 MHz with 1 channel. Calculations are done by considering only beacon enabled mode and with only one sender and receiver. However, it consumes high power. As number of sender increases, efficiency of 802.15.4 decreases. Throughput of 802.15.4 declines and delay increases when multiple radios are used because of increase in number of collisions.

A lot of work is done to improve the performance of IEEE 802.15.4 and many improvements are made in improving this standard, where very little work is done to find out performance of this standard by varying data rates and also considering ACKnowledgement (ACK) and no ACK condition and how it affects delay, throughput, end-to-end delay and load. We get motivation to find out the performance of this standard with parameters load, throughput, delay and end to end delay at varying data rates.

\section{Overview and Operation of IEEE 802.15.4}
IEEE 802.15.4 is proposed as standard for low data rate, low power Wireless Personal Area Networks (WPANs) [1],[2]. In WPANs, end nodes are connected to a central node called coordinator. Management, in-network processing and coordination are some of key operations performed by coordinator. The super-frame structure in beacon enabled mode is divided into active and inactive period. Active period is subdivided into three portions;  a beacon, Contention Access Period (CAP) and Contention Free Period (CFP). In CFP, end nodes communicate with central node (Coordinator) in dedicated time slots. However, CAP uses slotted CSMA/CA. In non-beacon enabled mode, IEEE 802.15.4 uses unslotted CSMA/CA with Clear Channel Assessment (CCA) for channel access.

In [2], IEEE 802.15.4 MAC protocol non-beacon enabled mode is used. Nodes use un-slotted CSMA/CA operation for channel access and packet transmission. Two main variables that are needed for channel access algorithm are Back off Exponent (BE) and Number of Back offs (NB). NB is the number of times CSMA/CA algorithm was required to back off while attempting channel access and BE is related to how many back off periods, node must wait before attempting channel access. Operation of CSMA/CA algorithm is defined in steps below:

$1.$ NB and BE initialization: First, NB and BE are initialized, NB is initialized to 0 and BE to macMinBE which is by default equal to 3.\\
$2.$ Random delay for collision avoidance: To avoid collision algorithm waits for a random amount of time randomly generated in range of $ 2^{BE}-1 $, one back off unit period is equal to $20 T_s$ with $T_s=16\mu$s\\
$3.$ Clear Channel Assessment: After this delay channel is sensed for the unit of time also called CCA. If the channel is sensed to be busy,  algorithm goes to step 4 if channel is idle algorithm goes to step 5.\\
$4.$ Busy Channel:  If channel is sensed busy then MAC sub layer will increment the values of BE and NB, by checking that BE is not larger than $BE_{max}$. If value of NB is less than or equal to $NB_{max}$, then CSMA/CA algorithm will move to step 2. If value of NB is greater than $NB_{max}$, then CSMA/CA algorithm will move to step 5  "Packet Drop", that shows the node does not succeed to access the channel.\\
$5.$ Idle Channel: If channel is sensed to be idle then algorithm will move to step 4 that is ``Packet Sent'', and data transmission will immediately start.

Fig. 1 illustrates aforementioned steps of CSMA/CA algorithm, starting with node has some data to send.

\begin{figure}[!h]
\centering
\includegraphics[width=3 in, height=4.5 in]{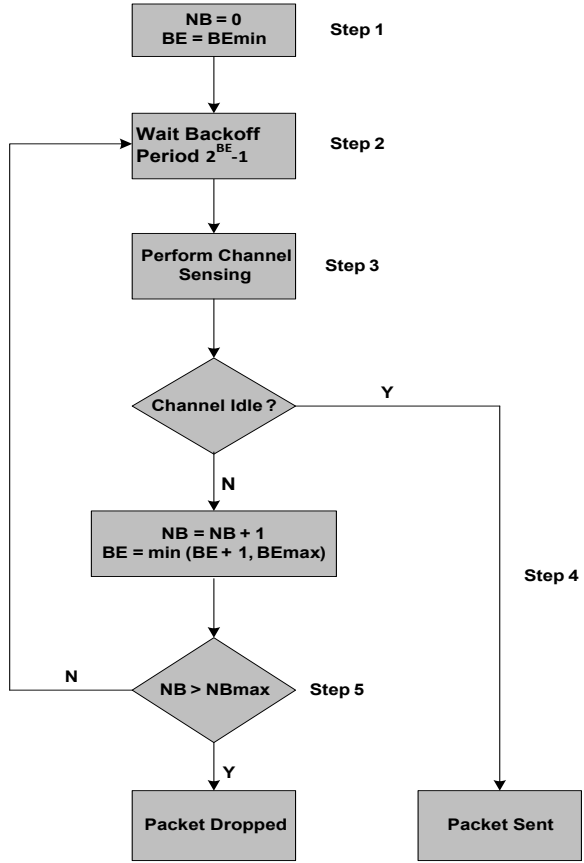}
\caption{Flow Chart of CSMA/CA Operation}
\end{figure}

\subsection{CSMA/CA}
CSMA/CA is a modification of Carrier Sense Multiple Access (CSMA). Collision avoidance is used to enhance performance of CSMA by not allowing node to send data if other nodes are transmitting. In normal CSMA nodes sense the medium if they find it free, then they transmits the packet without noticing that another node is already sending packet, this results in collision. CSMA/CA results in reduction of collision probability.

It works with principle of node sensing medium, if it finds medium to be free, then it sends packet to receiver. If medium is busy then node goes to backoff time slot for a random period of time and wait for medium to get free. With improved CSMA/CA, Request To Send (RTS)/Clear To Send (CTS) exchange technique, node sends RTS to receiver after sensing the medium and finding it free. After sending RTS, node waits for CTS message from receiver. After message is received, it starts transmission of data, if node does not receive CTS message then it goes to backoff time and wait for medium to get free. CSMA/CA is a layer 2 access method, used in 802.11 Wireless Local Area Network (WLAN) and other wireless communication. One of the problems with wireless data communication is that it is not possible to listen while sending, therefore collision detection is not possible.

CSMA/CA is largely based on the modulation technique of transmitting between nodes. CSMA/CA is combined with Direct Sequence Spread Spectrum (DSSS) which helps in improvement of throughput. When network load becomes very heavy then Frequency Hopping Spread Spectrum (FHSS) is used in congestion with CSMA/CA for higher throughput, however, when using FHSS and DSSS with CSMA/CA in real time applications then throughput remains considerably same for both. Fig. 2 shows the timing diagram of CSMA/CA.

\begin{figure}[!h]
\centering
\includegraphics[width=3.5 in, height=2 in]{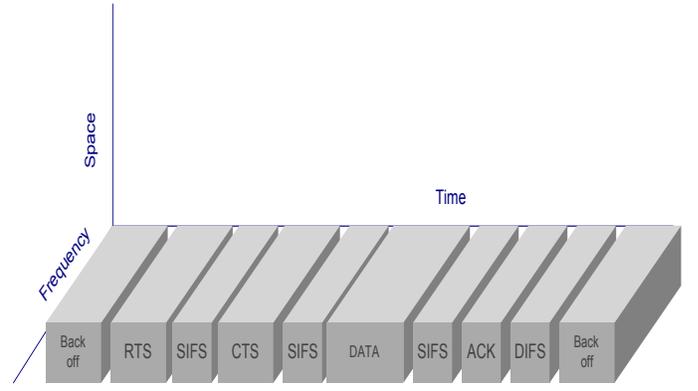}
\caption{Timing Diagram of CSMA/CA}
\end{figure}

\begin{eqnarray}
D=T_{bo}+T_{data}+T_{ta}+T_{ack}+T_{ifs}
\end{eqnarray}

Data transmission time $T_{data}$, Backoff slots time $T_{bo}$, Acknowledgement time $T_{ack}$ are given by equation 2, 3, and 4 respectively[2].
\\
\begin{table*}[htbp!]
  \centering
  \caption{Statistical Data of Delay, Throughput, Load and End to End delay of IEEE 802.15.4 at 20, 40, 250 Kbps}
    \begin{tabular}{|c|c|c|c||c|c|c||c|c|c||c|c|c|}
    \hline
    Time(minute)& \multicolumn{3}{|c||}{Delay(sec)} & \multicolumn{3}{|c||}{Throughput(bits/sec)} & \multicolumn{3}{|c||}{Load(bits/sec)} & \multicolumn{3}{|c|}{End to End delay(sec)}\\ \cline{2-13}
         & 20    & 40     & 250     & 20     & 40     & 250     & 20     & 40     & 250     & 20     & 40     & 250 \\
    \hline

    \hline
    0     & 9     & 2.5     & 0.00687     &  3511    & 7167     & 8347     & 8307     & 8307     & 8307     & 9     & 3     & 0.00521 \\
    \hline
    5     & 180     & 27     & 0.00686     &    4229  & 8709     & 10203     & 10190     & 10190     & 10190     & 93     & 30     & 0.00664 \\
    \hline
    10     & 374     & 54     & 0.00685     & 4296    & 8770     & 10293     & 10283     & 10283     & 10283     & 190     & 56     & 0.00675 \\
    \hline
    15    & 548     & 83   & 0.00685     & 4296     & 8778     & 10330     & 10319     & 10319     & 10319     & 277     & 83     & 0.00675 \\
    \hline
    20    & 721     & 111    & 0.00685     & 4313     & 8795     & 10349     & 10338     & 10338     & 10338     & 363     & 109     & 0.00675 \\
    \hline
    25     & 915    & 135   & 0.00684     & 4319    & 8802     & 10360     & 10351     & 10351     & 10351     & 461     & 136     & 0.00675 \\
    \hline
    30    & 1088    & 165     & 0.00683     & 4326    & 8808     & 10369    & 10358     & 10358     & 10358     & 547     & 165     & 0.00675 \\
    \hline
    35    & 1282     & 195     & 0.00683     & 4333     & 8804     & 10375     & 10364     & 10364     & 10364     & 634     & 195     & 0.00676 \\
    \hline
    40    & 1452     & 222     & 0.00683     & 4339     & 8799     & 10379     & 10369     & 10369     & 10369     & 731     & 221     & 0.00676 \\
    \hline
    45    & 1624     & 248     & 0.00683     & 4341     & 8801     & 10382     & 10372     & 10372    & 10372     & 817     & 248     & 0.00676 \\
    \hline
    50    & 1800     & 306     & 0.00683    & 4347     & 8807     & 10385     & 10375     & 10375     & 10375     & 903     & 278     & 0.00676 \\
    \hline
    55    & 1995     & 336     & 0.00683     & 4349     & 8805     & 10387     & 10377     & 10377     & 10377     & 1000     & 304     & 0.00676 \\
    \hline
    60    & 2145     & 380     & 0.00683     & 4352     & 8805     & 10388     & 10379     & 10379     & 10379     & 1076     & 328     & 0.00676 \\
    \hline
    \end{tabular}
  \label{tab:addlabel}
\end{table*}

\begin{eqnarray}
T_{data}=\frac{L_{phy} + L_{mac hdr} + payload + L_{mac ftr}}{R_{data}}\\
\nonumber\\
T_{bo}=bo_{slots} * T_{boslots}\\
\nonumber\\
T_{ack}= \frac{L_{phy} + L_{mac hdr} + L_{mac ftr}}{R_{data}}
\end{eqnarray}
The following notations are used:
\\\\
$T_{bo}=Back$ $Off$ $Period$
\\
$T_{data}=Transmission$ $Time$ $of$ $Data$
\\
$T_{ta}=Turn$ $Around$ $Time$
\\
$T_{ack}=Acknowledgement$ $Transmission$ $Time$
\\
$T_{ifs}=Inter$ $Frame$ $Space$
\\
$T_{phy}=Length$ $of$ $Physical$ $header$
\\
$L_{mac hdr}=Number$ $of$ $MAC$ $header$
\\
$Payload=Number$ $of$ $data$ $byte$ $in$ $packet$
\\
$L_{mac ftr}=Size$ $of$ $MAC$ $footer$
\\
$bo_{slots}=Number$ $of$ $back$ $off$ $slots$
\\
$T_{bo slots}= Time$ $for$ $a$ $back$ $off$ $slot$
\\
$R_{data}= Data$ $Rate$

In CSMA/CA mechanism, packet may loss due to collision. Collision occurs when two or more nodes transmits the data at the same time. If ACK time is not taken in to account then there will be no retransmission of packet and it will be considered that each packet has been delivered successfully.
The probability of end device successfully transmitting a packet is modeled as follows[3].

\begin{eqnarray}
P_{backoff period}=\frac{1}{2^{BE}}
\\
\nonumber\\
P_{tss}= \frac{1}{D}({1-\frac{1}{D}})^{BE-2} \\
= p{(1-p)}^{BE-2} \nonumber
\end{eqnarray}\newline where, $D$ is the number of end devices that are connected to router or coordinator. BE is the backoff exponent in our case it is 3. $P_{tss}$ is the probability of transmission success at a slot. $\frac{1}{D}$ is the probability of end device successfully allocated a wireless channel.

General formula for $P_{time delay event}$ is given by equation 8. Probability of time delay caused by CSMA/CA backoff exponent is estimated as in [7]. Maximum number of backoff is 4. Value of BE=3 has been used in following estimation and we estimate by applying summation from 3 to 5. $P_{tde}$ is the probability of time delay event.

\begin{eqnarray}
P_{tde}= \sum_{n=0}^{2^{BE-1}} n\frac{1}{2_{BE}}p{1-p}^{BE-2}
\end{eqnarray}
\begin{eqnarray}
P_{tde}= \sum_{n=0}^{7} n\frac{1}{2_{BE}}p + \sum _{n=8}^{15} n\frac{1}{2_{BE}}p + \sum _{n=16}^{31} n\frac{1}{2_{BE}}p
\end{eqnarray}
Expectation of the time delay is obtained as from [7]. $P{E_{A}}$ and $P{E_{B}}$ are taken from equations 7 and 8 respectively.
\begin{eqnarray}
E[Time Delay]=P({E_{A}|E_{B}})\nonumber\\
\nonumber\\
= \frac{\sum_{n=0}^{7} n\frac{1}{2_{BE}}p + \sum _{n=8}^{15} n\frac{1}{2_{BE}}p + \sum _{n=16}^{31} n\frac{1}{2_{BE}}p}{\sum_{n=0}^{2^{BE-1}} n\frac{1}{2_{BE}}p{1-p}^{BE-2}}
\end{eqnarray}
\subsection{Analysis of Statistical and Simulation Data of IEEE 802.15.4}
Statistical data of throughput, load, end-to-end delay and delay of IEEE 802.15.4 at varying data rates is shown in table I. It shows different values of delay, throughput, end-to-end delay and load recorded at different time. Load at all data rates and at all time intervals remains same. Start time for simulation is kept at 0 seconds and stop time is kept to infinity. Load in all three data rates at different time intervals remains same as shown in table I. There is very small difference between delay and end-to-end delay. At 20 Kbps maximum delay of 2145 seconds is recorded with maximum throughput of 4352 bits/sec at 60 min. At 40kbps maximum delay of 380 seconds and minimum delay 2.5 seconds is recorded. Throughput of 8805 (bits/sec) is the highest throughput recorded on 60 min. In case of 250 Kbps delay remains very small, near to negligible where as throughput matches load with 10388 (bits/sec).

\begin{table}
\caption {Network Parameters of 802.15.4}
\begin {center}
\begin {tabular} {| p{2.5cm} | p{2cm} |}
\hline
Parameter type & Value \\ \hline
Beacon order &   6 \\ \hline
Superframe order & 0 \\ \hline
Maximum routers & 5 \\ \hline
Maximum depth & 5 \\ \hline
Beacon Enabled Network & Disabled \\ \hline
Route Discovery Time  & 10(sec) \\ \hline
\end{tabular}
\end{center}
\end{table}

Network parameter are given in table II. Non beacon mode is selected in our analysis and beacon order is kept at 6. Due to non-beacon enabled mode superframe order is not selected. Maximum routers or nodes that can take part in simulation is 5, each having tree depth of 5. Discovery time that is needed by each router to discover route is 10 sec.

\begin{table}[h]
\caption {Simulation Parameters of 802.15.4}
\begin {center}
\begin {tabular} {| p{3cm} | p{2cm} |}
\hline
Parameter type & Value \\ \hline
Minimum Backoff exponent &   3 \\ \hline
Maximum number of backoff & 5 \\ \hline
Channel sensing duration & 0.1 (sec) \\ \hline
Data Rates & 20, 40, 250 kbps \\ \hline
Packet reception power & -85 (dbm) \\ \hline
Transmission band  & 2.4 (MHz) \\ \hline
Packet size & 114 bytes \\ \hline
Packet interarrival time  & 0.045(sec) \\ \hline
Transmit Power & 0.05 (W)\\ \hline
ACK wait duration & 0.05 (sec) \\ \hline
Number of retransmissions & 5 \\ \hline
\end{tabular}
\end{center}
\end{table}

\section{Simulation study of IEEE 802.15.4}

Simulation parameters of 802.15.4 with its value are shown in table III. Minimum BE is kept at 3 with maximum no. of back-off to 5. Default settings of 802.15.4 are used in this simulation. Packet reception power is kept at -85 dbm with transmitting power of 0.5 watt(W). In ACK enabled case, ACK wait duration is kept at 0.05 sec with no of retransmissions to 5. In no ACK case these parameters are disabled. 114 bytes is the packet size with interarrival time of 0.045 sec. Transmission band used in this simulation is 2.4 GHz. Simulations have been performed at varying data rates of 20, 40, 250 kbps.

\begin{figure}[t]
\centering
\includegraphics[width=3.5 in, height=2.5 in]{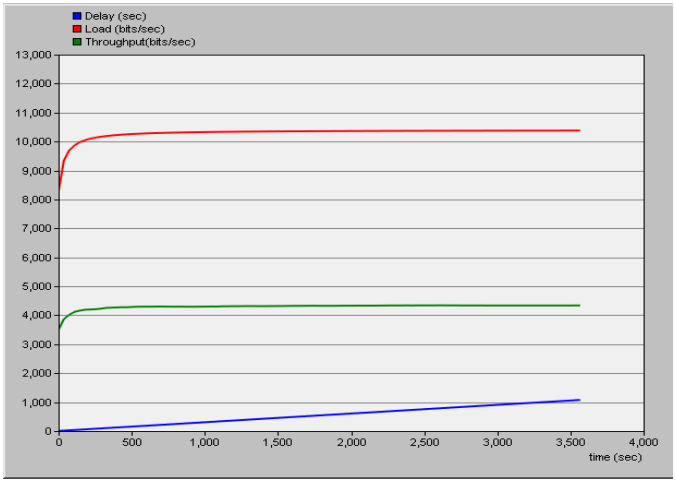}
\caption{Delay, Throughput and Load at 20 Kbps}
\end{figure}

\begin{figure}[t]
\centering
\includegraphics[width=3.5 in, height=2.5 in]{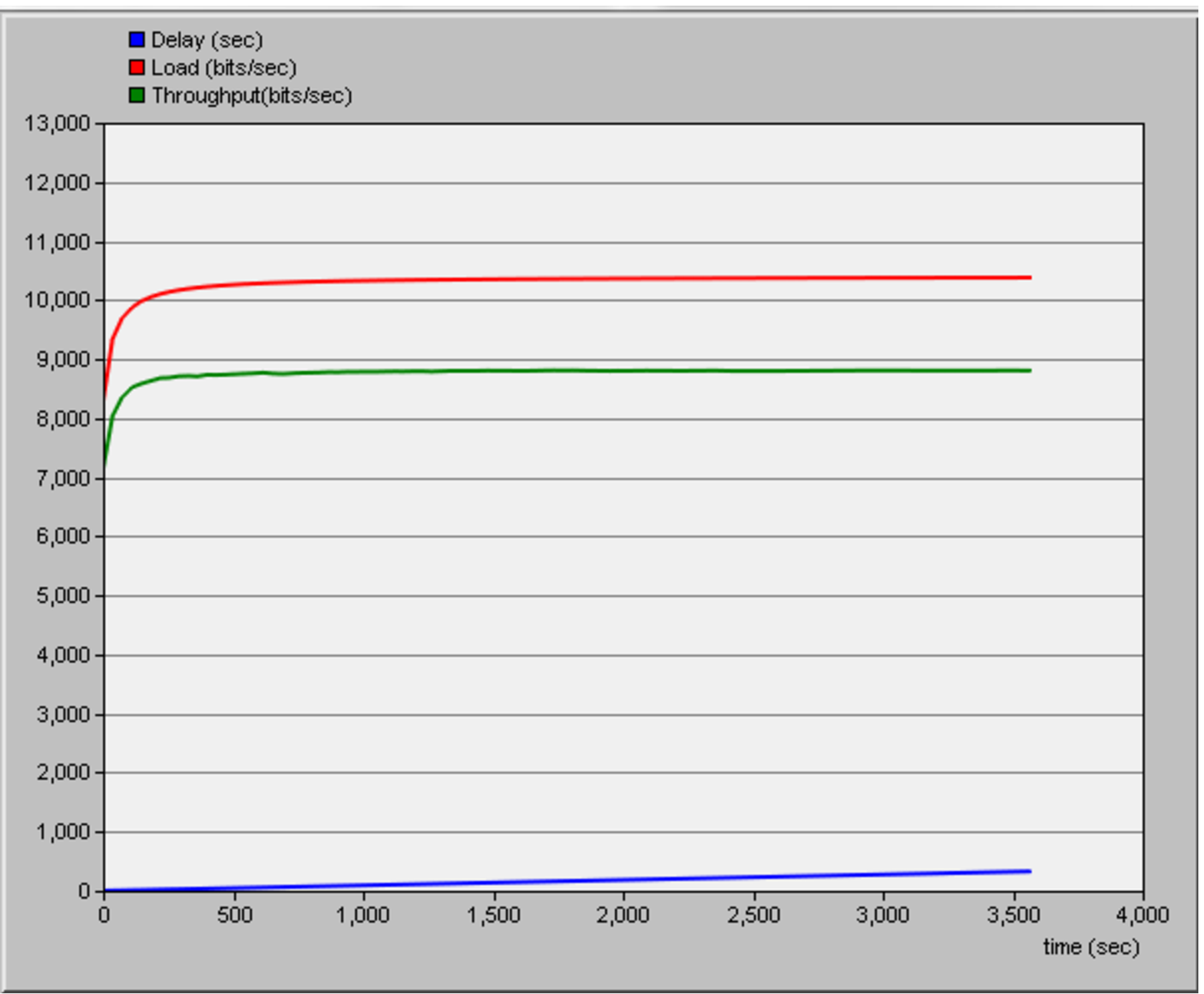}
\caption{Delay, Throughput and Load at 40 Kbps}
\end{figure}

\begin{figure}[h]
\centering
\includegraphics[width=3.5 in, height=2.5 in]{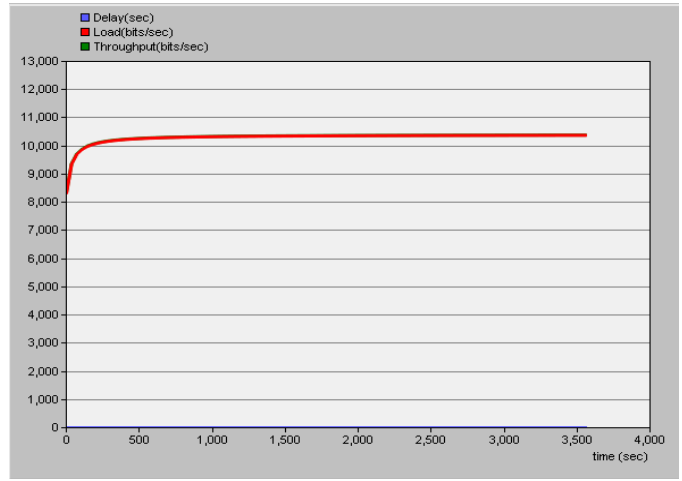}
\caption{Delay, Throughput and Load at 250 Kbps}
\end{figure}

Simulations for both ACK and non ACK cases have also been performed. OPNET modeler is the simulator used for simulations. Simulations are executed for one hour with update interval of 5000 events. Graphs are presented in overlaid statistics. Overlaid means that, graphs of each scenario has been combined with each other. Data of graphs are averaged over time for better results. Personal Area Network IDentity (PAN ID) is kept at default settings, coordinator automatically assigns PAN ID to different personal area networks if they are attached. We consider non beacon mode for our simulations. Using non-beacon enabled mode improves the performance and changing different parameters affects performance of 802.15.4. CSMA/CA values are kept to default with minimum backoff exponent to 3 and having maximum backoff of 5. Changing these parameters does not affect its performance. We perform simulations with ACK and non ACK. In non ACK there is only delay due to node waiting while sensing medium, there is no delay due to ACK colliding with packets. In ACK case there is collision for packets going towards receiver and ACK packet coming from receiver at same time. Delay in ACK is more as compare to non ACK case. We use standard structure of IEEE 802.15.4 with parameters shown in table II.
%f6
\begin{figure}[t]
\centering
\includegraphics[width=3.5 in, height=2.5in]{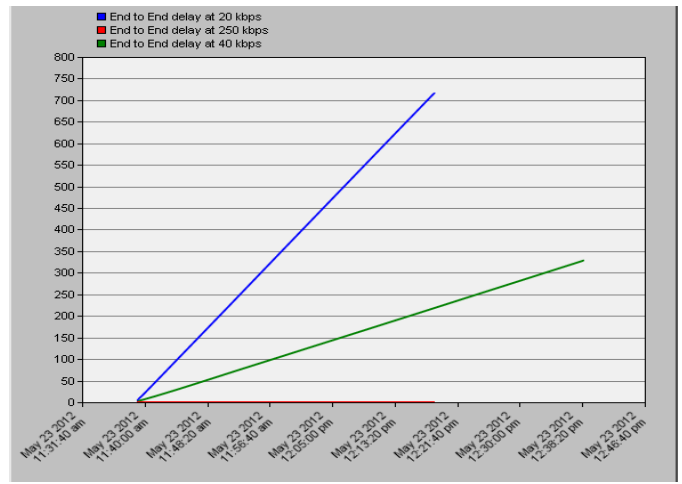}
\caption{End to End Delay at 20, 40, 250 Kbps}
\end{figure}

%f7
\begin{figure}[t]
\centering
\includegraphics[width=3.5 in, height=2.5in]{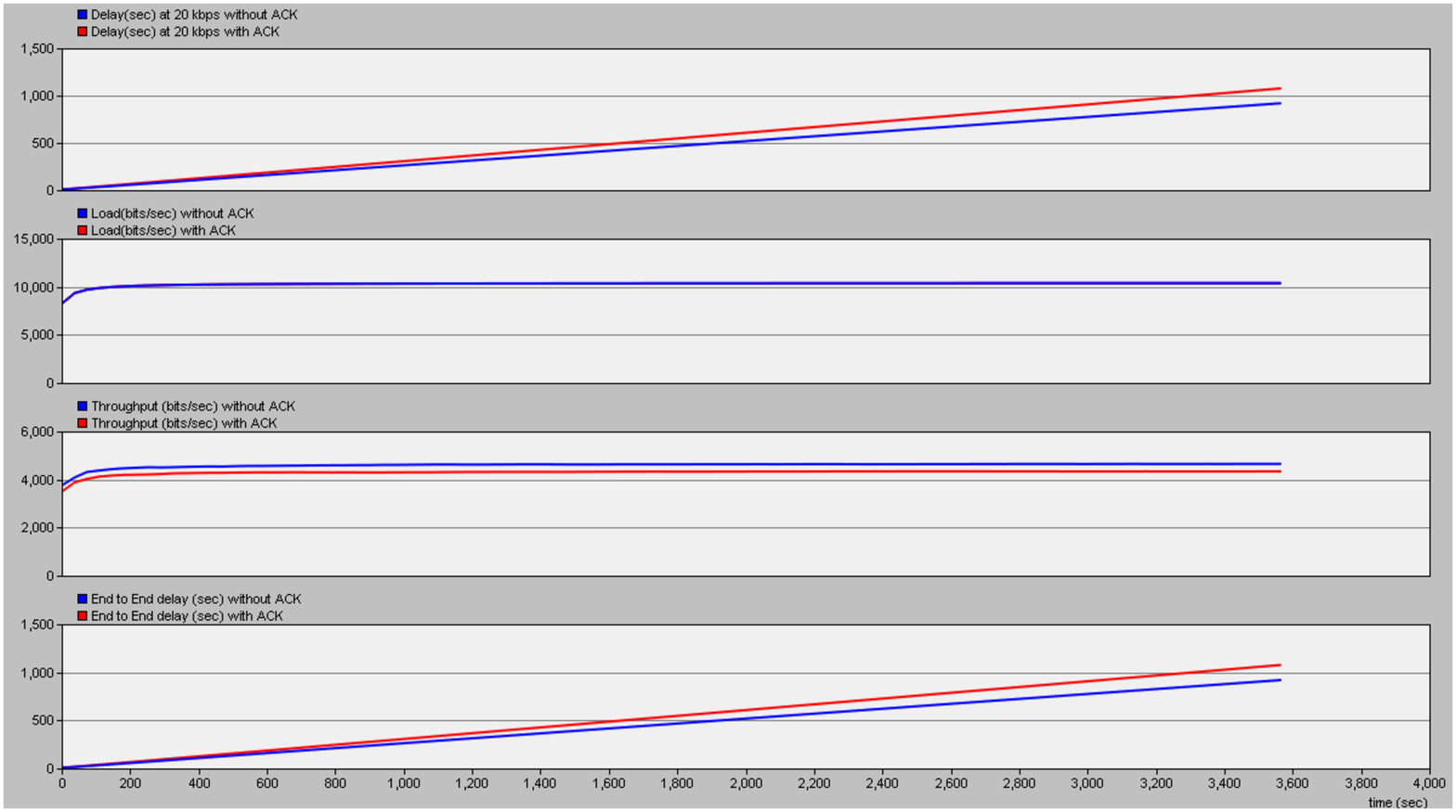}
\vspace{-0.4cm}
\caption{Delay, Throughput, Load and End to End delay, with and without ACK at 20 Kbps}
\end{figure}

%f8
\begin{figure}[t]
\centering
\includegraphics[width=3.5 in, height=2.5in]{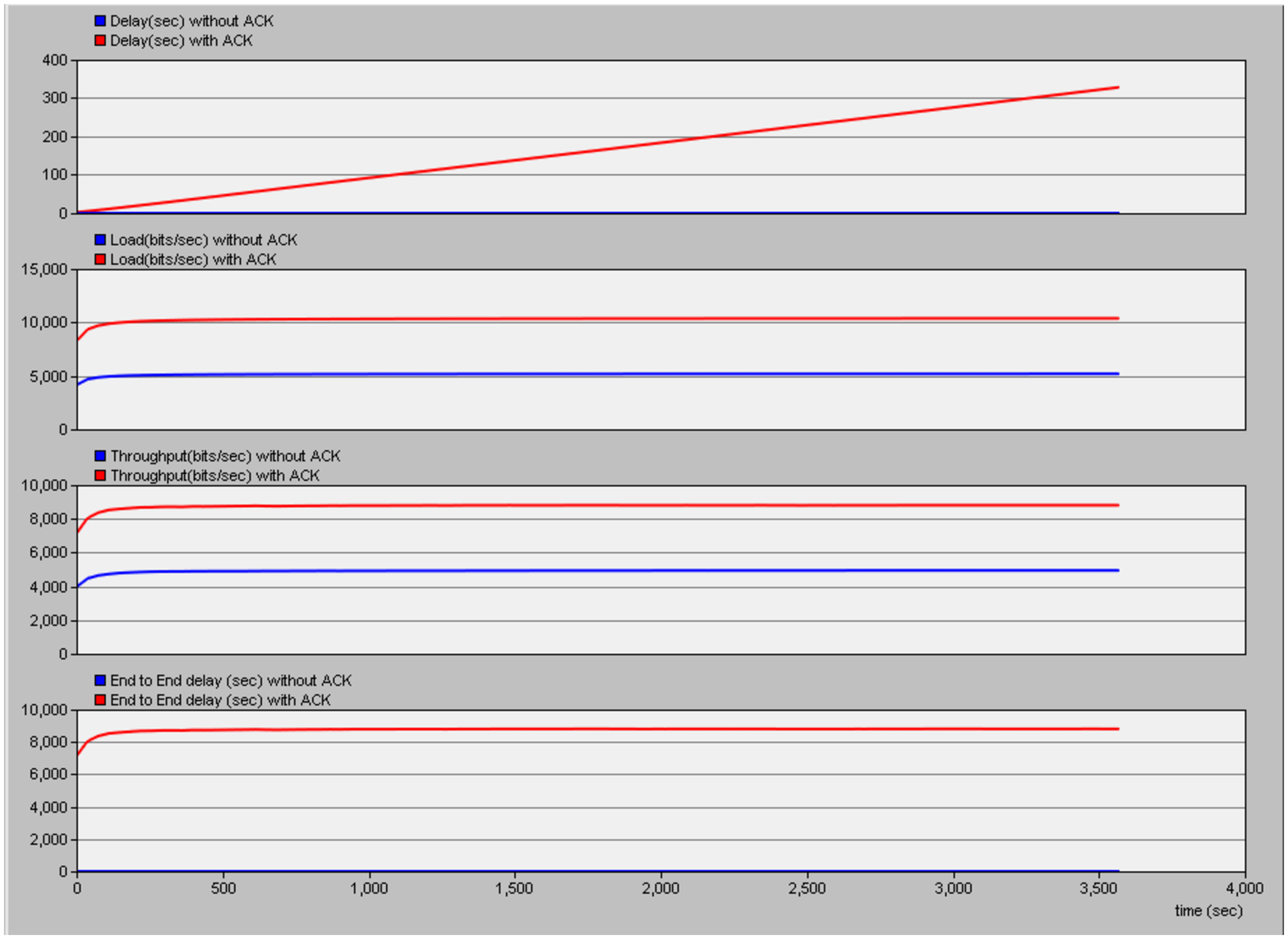}
\vspace{-0.4cm}
\caption{Delay, Throughput, Load and End to End delay, with and without ACK at 40 Kbps}
\end{figure}

%f9
\begin{figure}[b]
\centering
\includegraphics[width=3.5 in, height=2.5in]{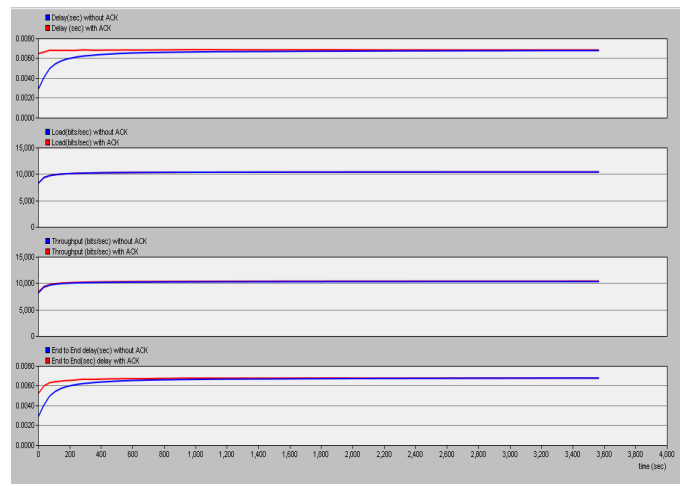}
\vspace{-0.4cm}
\caption{Delay, Throughput, Load and End to End delay, with and without ACK at 250 Kbps}
\end{figure}

In this section, performance of default MAC parameters of IEEE 802.15.4 standard non beacon enabled mode. Simulations are performed considering 10 sensor nodes environment with coordinator collecting data from all nodes. Fig 3, 4, 5 and 6 show graphical representation of performance parameters of 802.15.4.

Delay represents the end-to-end delay of all the packets received by 802.15.4 MACs of all WPANs nodes in the network and forwarded to the higher layer.

Load represents the total load in (bits/sec) submitted to 802.15.4 MAC by all higher layers in all WPANs nodes of the network. Load remains same for all the data rates. Throughput represents the total number of bits in (bits/sec) forwarded from 802.15.4 MAC to higher layers in all WPANs nodes to the network. End-to-end delay is the total delay between creation and reception of an application packet.

Delay, load and throughput are plotted as function of time. As load is increasing, there is increase in throughput and delay. When load becomes constant, throughput also becomes constant, however, delay keeps on increasing. Delay in 802.15.4 occurs due to collision of data packets or sometimes nodes keeps on sensing channel and does not find it free. When node senses medium and find it free, it sends packet. At same time, some other nodes are also sensing the medium and find it free, they also send data packets and thus results in collision. Collision also occurs due to node sending data packet and at same time coordinator sending ACK of successfully receiving packet and causing collision. When ACK is disabled this type of collision will not occur.

Delay, throughput and load is analyzed at 40 Kbps in Fig 4. With increase in load, there is increase in throughput and delay, however, it is less as compared to 20kbps, this is due to increase in data rate of 802.15.4. Increase in bit transfer rate from 20 to 40kbps causes decrease in delay and hence increases throughput.

Fig 5 shows behavior of 802.15.4 load, throughput and delay at 250kbps data rate. Delay is negligible at this data rate, with throughput and load showing same behavior. Delay approaching zero shows that, at 250 Kbps data rate there are less chances of collision or channel access failure. IEEE 802.15.4 performs best at this data rate compared to 20 and 40 Kbps.

At same time end-to-end delay of IEEE 802.15.4 at varying data rates of 20, 40 and 250 Kbps are shown in Fig 6. This figure shows that end to end delay for 20 Kbps data rate is higher than 40 Kbps and 250 Kbps. Minimum end-to-end delay is found at 250 Kbps data rate. At 250 Kbps, more data can pass at same time with less collision probability hence having minimum delay and at 20 Kbps, less data transfers at same time causing more end to end delay. Statistical data of end-to-end delay is shown in table I, which shows end to end variation with change in time.

Fig 7 shows the delay, throughput, load and end-to-end delay of IEEE 802.11.4 at 20 Kbps data rate with and without ACK. Load remains same in both cases. There is no collision because of ACK packets due to which packets once send are not sent again. There is decrease in delay and increase in ACK due to less collision.

End-to-end delay performs same as delay. IEEE 802.15.4 performs better with non ACK other than ACK due to decrease in collision probability in no ACK compared to ACK case.

Delay, throughput, load and end-to-end delay with and without ACK at 40 Kbps are presented in Fig 8. There is considerable difference between the analysis in ACK and without ACK case. Delay is reduced to negligible at low value of $0.045$ in no ACK case due to reason that, at this data rate there is no collision therefor, delay is nearly zero. As there is no collision and channel sensing time is also low, this increase throughput and load in non ACK case, as compared to ACK.

Fig. 9 shows analysis with ACK and no ACK cases of delay, throughput, load and end to end delay at 250 Kbps, at  this high data rates load and throughput in both cases becomes equal to each other and data is sent in first instant to coordinator by nodes. Delay in both cases nearly equal to zero, which shows that, there is very less collision at this high data rates and channel sensing time is also very low. End to End delay slightly differs from delay in no ACK case.

\section{Conclusion and Future Work}
In this paper, performance of IEEE 802.15.4 standard with non-beacon enabled mode is analyzed at varying data rates. We have evaluated this standard in terms of load, delay, throughput and end-to-end delay with different MAC parameters. We have also analyzed performance with ACK enabled mode and no ACK mode. We considered a full size MAC packet with payload size of 114 bytes for data rates 20 Kbps, 40 Kbps and 250 Kbps. It is shown that better performance in terms of throughput, delay, and end-to-end delay is achieved at higher data rate of 250kbps. IEEE 802.15.4 performs worse at low data rates of 20kbps. Performance of this standard improves with increase in data rate.
\newpage
We have shown in our paper through statical and graphical data that performance of standard IEEE 802.15.4 improves with increase in data rates and with decrease in data rate its performance degrades.

In future research work, we will investigate the performance of IEEE 802.15.4 in WBANs by changing frequency bands on different data rates. We also intend to examine the effect of changing inner structure of MAC layer in IEEE 802.15.4.


\begin{thebibliography}{1}
\bibitem{Reference 1}F. Timmons, N $et$ $al$., ``Analysis of the Performance of IEEE 802.15.4 for Medical Sensor Body Area Networking'', Sensor and Ad Hoc Communications and Networks, 2004.
\bibitem{Reference 2}C. Buratti and R. Verdone $et$ $al$., ``Performance Analysis of IEEE 802.15.4 Non Beacon-Enabled Mode'', IEEE Transaction on vehicular technology, Vol. 58, No. 7, September 2009.
\bibitem{Reference 3}Anastasi, G $et$ $al$., ``The MAC Unreliability Problem in IEEE 802.15.4 Wireless Sensor Networks'', MSWiM09 Proceedings of the 12th ACM international conference on Modeling, analysis and simulation of wireless and mobile systems, October 2009.
\bibitem{Reference 4}S. Ullah, K. S. Kwak $et$ $al$., ``An Ultra-Low power and Traffic-Adaptive Medium Access Control Protocol for Wireless Body Area Network'', J Med Syst, DOI 10.1007/s10916-010-9564-2.
\bibitem{Reference 5}S. Ullah, B.Shen, S.M.R. Islam, P. Khan, S. Saleem and K.S. Kwak $et$ $al$., ``A Study of Medium Access Control Protocols for Wireless Body Area Networks''.
\bibitem{Reference 6}X. Liang and I. Balasingham $et$ $al$., ``Performance Analysis of the IEEE 802.15.4 based ECG Monitoring Network''.
\bibitem{Reference 7}B. Latre, P.D. Mil, I. Moerman, B. Dhoedt and P. Demeester $et$ $al$., ``Throughput and Delay Analysis of Unslotted IEEE 802.15.4'',  Journal of Networks, Vol. 1, No. 1, May 2006.
\end{thebibliography}
\end{document}